\documentclass{article}

\usepackage[margin=1in]{geometry}
\usepackage{hyperref}
\usepackage{parskip}
\usepackage{setspace}
\usepackage{bm}
\usepackage{amsmath}
\usepackage{float}
\usepackage{adjustbox}
\usepackage{graphicx} 
\usepackage{natbib}

\title{A Bayesian Semiparametric Mixture Model for Clustering Zero-Inflated Microbiome Data}
\author{Suppapat Korsurat\footnote{Department of Statistics, Colorado State University, Fort Collins, CO, USA }  { and} Matthew D. Koslovsky\footnote{Department of Statistics, Colorado State University, Fort Collins, CO, USA, \href{mailto:matt.koslovsky@colostate.edu}{matt.koslovsky@colostate.edu}} }
\date{}

\begin{document}

\maketitle

\noindent \textbf{Abstract.} Microbiome research has immense potential for unlocking insights into human health and disease.  A common goal in human microbiome research is identifying   subgroups of individuals with similar microbial composition that may be linked to specific health states or environmental exposures. However, existing clustering methods are often not equipped to  accommodate the complex structure of microbiome data and typically make limiting assumptions regarding the number of clusters in the data  which can bias inference. Designed for zero-inflated multivariate compositional count data collected in microbiome research, we propose a novel Bayesian semiparametric mixture modeling framework that  simultaneously learns the number of clusters in the data while performing cluster allocation.  In simulation, we demonstrate the clustering performance of our method compared to distance- and model-based alternatives and the importance of accommodating zero-inflation when present in the data. We then  apply the model to identify clusters in microbiome data collected in a study designed to investigate the relation between gut microbial composition and enteric diarrheal disease.




\section{Introduction}

 The human microbiome is a complex ecosystem of microorganisms residing within and on our bodies. Each individual possesses a unique microbiome pattern which is  influenced by various external factors, such as environment, climate, diet, and medical conditions \citep{ursell2012defining, rojo2017exploring, allaband2019microbiome, ahn2021environmental}. 
 A common goal in human microbiome research is identifying   subgroups of individuals with similar microbial composition, referred to as enterotypes, that may be linked to specific health states or environmental exposures \citep{arumugam2011enterotypes,holmes2012dirichlet, marcos2021applications}.
  For example, this research was motivated by data collected in a study investigating the relation between intestinal microbial community composition and enteric infection \citep{singh2015intestinal}.  Given the critical role  intestinal microbiota play in maintaining a healthy immune response, there is considerable interest in uncovering patterns in microbial composition to  investigate the feasibility of microbiota-based diagnostics, therapies, or prevention of disease, potentially through personalized treatment strategies  \citep{sekirov2009role,costea2018enterotypes}.   

\vspace{ 0.25cm }

  Typically, microbiome data take the form of an $N \times J$-dimensional matrix of counts, where $N$ represents the number of observations and $J$ represents the number of unique microbial taxa.  The conventional approach for obtaining taxa counts is to sequence the 16S rRNA gene, as it contains well-conserved and hypervariable regions to differentiate different species. Then, sequenced reads are clustered into operational taxonomic units, or OTUs, and classified to a reference database using various methods \citep{huson2007megan,wang2007naive, edgar2013uparse,  allard2015spingo, bolyen2019reproducible}. More recently, researchers have promoted the use of amplicon sequence variants (ASVs) instead of OTUs as the unit of analysis in microbiome research, which can distinguish sequence variants differing by as little as one nucleotide \citep{eren2013oligotyping,callahan2016dada2,amir2017deblur,callahan2017exact}. Regardless of the approach taken, these data are inherently challenging to analyze due to their high-dimensionality, overdispersion, compositional structure, and  zero-inflation.
  
  Zero reads in microbiome data occur when (1) the organism is not present in the sampling region, and therefore the probability of occurrence is zero (i.e., structural zero), or (2) the organism is present but was not sampled (i.e., at-risk zero). To model zero-inflation, researchers typical construct a two-component mixture of a point mass at zero and a sampling distribution for the counts where a latent at-risk indicator is introduced to differentiate between at-risk and structural zeros \citep{ xu2015assessment, neelon2019bayesian, zhang2020nbzimm,shuler2021bayesian, jiang2023flexible, koslovsky2023bayesian, koslovsky2025analyzing}. While numerous methods have been developed to accommodate zero-inflation in microbiome research, including network analysis \citep{ha2020compositional}, dimension reduction \citep{zeng2021model, koslovsky2023bayesian}, regression modeling \citep{tang2019zero}, longitudinal data analysis \citep{zhang2020nbzimm, zhang2020zero}, association tests \citep{ling2021powerful}, and causal inference \citep{yang2023estimation}, among others, existing methods for cluster analysis often are not equipped to accommodate the complex structure of the data and may make limiting assumptions regarding the number of clusters in the data which can bias inference. 

Two common approaches for identifying clusters of microbial samples are distance-based methods (e.g., K-means, partitioning around the median (PAM), and hierarchical clustering \citep{ xu2015comprehensive,  shi2022performance}) and model-based methods \citep{holmes2012dirichlet, subedi2020cluster, mao2022dirichlet, shi2023sparse}. Distance-based methods use the distance between two observations to determine cluster allocation without assuming a statistical distribution for the observed counts. A key challenge in applying distance-based clustering methods  to microbiome data is choosing the proper distance metric to capture the similarity or dissimilarity of microbial communities, often referred to as  $\beta$-diversity \citep{namkung2020machine}. Several $\beta$-diversity metrics have been proposed, including Aitchison, Bray-Curtis, and UniFrac distances \citep{aitchison2000logratio, chen2012associating}. See \cite{Plantinga2021} for a detailed review of common $\beta$-diversity metrics used in microbiome data analysis  and  \cite{shi2022performance} for an extensive simulation study investigating  how clustering performance can depend on the chosen $\beta$-diversity metric and clustering algorithm. 

Model-based clustering methods are a popular alternative to distance-based methods that assume the population consists of observations from distinct clusters, each with their own statistical distribution. Finite mixture models (FMMs), which belong to the class of model-based clustering methods, assume that the population consists of observations originating from a finite number of underlying clusters. In mixture modeling, the concept of determining the number of clusters in the data lies in the distinction between $ K$, the number of components in the model (i.e., the number of potential clusters), and $K_{+}$, the number of clusters that are actually present in the data (i.e., non-empty components) \citep{miller2018mixture}. Typically the number of clusters in FMMs is specified prior to analysis (i.e., $K_+ = K$),  and model comparison techniques are used to select $K$ \citep{yeung2001model, marin2005bayesian}.   Alternatively, mixture of finite mixtures (MFM) models place a prior on the number of mixture components to draw inference on the number of clusters in the data \citep{miller2018mixture}. 
To relax assumptions on the number of clusters in the data, researchers commonly use infinite mixture models (IMMs), such as Dirichlet process mixture models (DPMMs), that allow $K = \infty$ \citep{mcauliffe2006nonparametric, li2019tutorial}.  While IMMs do not require specifying the number of clusters a priori, they have been shown to overestimate the true number of clusters, often leading to the formation of numerous singleton clusters \citep{miller2013simple, li2019tutorial,  fruhwirth2021generalized}.  Notably, \cite{ascolani2023clustering} recently showed the posterior for the number of clusters in DPMMs is consistent under certain conditions for the prior on the concentration parameter.
 
   \textit{Sparse} finite mixture models (sFMMs) were recently introduced as a semiparametric alternative for model-based clustering that bridges standard FMMs and IMMs and is recommended when the number of clusters is not expected to increase with larger sample sizes \citep{malsiner2016model,fruhwirth2019here}.  This approach deliberately overspecifies the number of potential components in the model (i.e., $K > K_+$) and then places a sparsity-inducing prior on the mixture weights that shrinks them  towards zero to encourage empty components. One of the challenges of implementing sFMMs is that clustering performance  has been shown to depend heavily on the chosen value of the mixture weights, or prior thereof \citep{celeux2019model,fruhwirth2019here}, and none of the empty components' probabilities are set exactly to zero. 

 In this work, we develop a Bayesian model-based clustering method for zero-inflated microbiome data. Our approach belongs to the class of MFM models and can be thought of as a \textit{discrete} alternative to existing sFMMs that similarly overspecifies the number of potential clusters in the data. However, instead of shrinking empty components' mixture weights towards zero, the proposed method  places a point mass at zero to   remove empty components from the model. Commonly, the Dirichlet-multinomial (DM) distribution and its extensions are used to model microbial counts and their corresponding relative abundances as it inherently accommodates the  compositional structure of the data and overdispersion \citep{holmes2012dirichlet,wadsworth2017integrative,harrison2020dirichlet}. In this work, we model the  multivariate count data  collected in the application study with a  zero-inflated DM (ZIDM) model  which further accounts for excess zeros typically found in microbiome data.  As such, our modeling approach extends the work of \cite{koslovsky2023bayesian} to account for heterogeneity in zero-inflated multivariate count data, effectively using a ZIDM distribution to model zero-inflation in the count data as well as induce sparsity in the mixture weights. Together, our approach accommodates the complex structure of microbiome data observed in practice, simultaneously estimates cluster-specific taxa relative abundances while performing cluster allocation, and does not require specifying the number of clusters in the data a priori.  For inference, we take a fully Bayesian approach and implement a telescoping sampler \citep{fruhwirth2021generalized} to help improve the mixing of the Markov chain Monte Carlo (MCMC) algorithm. We demonstrate how the proposed method outperforms alternative approaches for clustering microbiome data in a variety of simulation scenarios. In the application study, we identify two main clusters of microbial composition; one dominated by a combination of $Bacteroides$  and $Cronobacter$ and composed mostly of enteric diarrheal disease (EDD) patients, and  another dominated by $Bacteroides$ with a balance between healthy individuals and EDD patients.

\section{Methods}\label{sec::methods} 

 In this section, we first introduce standard notation and definitions used for model-based cluster analysis in the context of
 multivariate count data. Thereafter, we propose a novel discrete sparse Dirichlet-multinomial mixture model (DSDM$^3$) that we use to simultaneously accommodate uncertainty in the number of  clusters in the  data while performing cluster allocation. 
Let $\bm{z}_{i}$ represent the  $J$-dimensional vector of taxa counts for the $i^{th}$ individual, $i = 1, 2, {\dots}, N$. We assume 
\begin{equation}
    \bm{z}_{i} \mid c_{i} = k, \bm{\Theta}_{k} \sim F\left(\bm{\Theta}_{k}\right),
\end{equation}
\noindent where $c_{i} = k$  indicates the $i^{th}$ individual is assigned to the $k^{th}$, $k = 1, 2, \dots,  K $, cluster, $ K $ represents the number of components, $F\left(\cdot\right)$ is the assumed probability mass function for the multivariate count data, and $\bm{\Theta}_{k}$ is a multivariate set of parameters for the $k^{th}$ component, both of which we describe in more detail below.   Equivalently, we can formulate a mixture model as $f\left(\bm{z}_{i}\right) = \sum_{k=1}^{K}w_{k}F\left(\bm{\Theta}_{k}\right)$, where $w_{k}$  are the mixture weights that sum to one  with $\boldsymbol{w} = (w_1, \dots, w_K)$. 
It is then common to assume $\boldsymbol{w} \mid K \sim \mbox{Dirichlet}(\psi_1, \dots, \psi_K)$, where   $\psi_k \equiv \psi$, for $k=1,\dots,K$. Unlike FMMs in which $K$ is fixed a priori, MFM modeling frameworks place a prior distribution on the total number of cluster components. Common prior assumptions for $K$ include the truncated Poisson and more recently a beta-negative-binomial, which generalizes the Poisson, negative-binomial, and geometric distributions \citep{miller2018mixture,fruhwirth2021generalized}. 

\subsection{A Semiparametric  Mixture Model}
 
We propose a DSDM$^{3}$ that deliberately overspecifies the number of potential components and induces sparsity in the number of  components by allowing empty components' corresponding mixture weights to take on a zero value. As such, our approach draws similarities to sFMMs that shrink mixture weights \textit{towards} zero and also belongs to the class of MFM models, which we show below. Specifically, we assume the $i^{th}$, $i=1,\dots,N$, individual's cluster assignment 
\begin{align}\label{eq:cluster}
\begin{split}
    c_{i} \mid \bm{w} &\sim \text{Multinomial}\left(1, \bm{w}\right), \\
    w_{k} &= \frac{\psi_{k}}{\sum_{k'=1}^{K_{m}}\psi_{k'}}, \\
    \psi_{k} \mid \lambda_{k} & \sim \lambda_{k}\text{Gamma}\left(\theta, 1\right) + \left(1 - \lambda_{k}\right)\delta_{0}\left(\psi_{k}\right) \text{, and } \\
    \lambda_{k} &\sim \text{Bernoulli}\left(\pi_{\lambda}\right),
    \end{split}
\end{align}  
with the constraint that at least one $\lambda_{k} = 1$, $k=1,\dots,K_{m}$, where $K_{m}$ is the maximum  number of potential components, $\lambda_{k} \in \left\{0, 1\right\}$ indicates whether or not the $k^{th}$ component exists in the model, $\theta$ is a hyperparameter controlling the mixture weights for active components, $\delta_{0}\left(\cdot\right)$ is a Dirac delta function, or point mass, at zero, and the hyperparameter $\pi_{\lambda}$ represents the probability that the $k^{th}$ component exists in the model.  This formulation ensures that the probability of being assigned to the $k^{th}$ cluster is zero when $\lambda_{k}$ equals zero.  Ideally, $K_{m}$ should be set large enough to accommodate the true number of clusters in the data but not so large that it leads to excessive computations. 

As mentioned previously, the  DSDM$^3$ belongs to the class of MFM models. To make this connection, we first note that given the set of $\lambda_k = 1$, Equation \ref{eq:cluster} is equivalent to the standard Dirichlet-multinomial formulation used in mixture models with $K = \sum_{k^{\prime} =1}^{K_{m}} \lambda_{k^{\prime}}$ components. Then under the assumptions for $\lambda_k$ described above, the implied distribution for $K$ is a zero-truncated binomial distribution with $p\left(K = k\right) = \frac{\binom{K_{m}}{k} \pi_{\lambda}^{k} \left(1 - \pi_{\lambda}\right)^{K_{m} - k}}{1 - \left(1 - \pi_{\lambda}\right)^{K_{m}}}$. 
In the Supplementary Material, we provide more details of this connection, derivations for the induced distribution on $K_+$, the number of active clusters in the model  (i.e., the number of non-empty components), and the corresponding exchangeable partition probability function.  

\subsection{Zero-Inflated Multivariate Count Data}

In this work, we apply the proposed method to cluster zero-inflated multivariate compositional count data collected in human microbiome research.   The DM distribution is commonly used to model microbial counts as it accommodates the compositional structure of the data and overdispersion \citep{wadsworth2017integrative,harrison2020dirichlet, koslovsky2020microbiome, pedone2023subject,shi2023sparse}. However, naively assuming a DM for microbial counts can bias parameter estimates as it is not inherently equipped to handle zero-inflation \citep{koslovsky2023bayesian}. To accommodate potential zero-inflation, we assume  a ZIDM model for the multivariate compositional counts. Specifically, we let
\begin{align}
\begin{split}
    \bm{z}_{i} \mid  \bm{\phi}_{i} &\sim \text{Multinomial}\left(\sum_{j=1}^{J}z_{ij},   \frac{\boldsymbol{\phi}_{i}}{\sum_{j'=1}^{J}\phi_{ij'}}\right),\\ 
    \phi_{ij} \mid c_{i} = k, \gamma_{ij}, \xi_{kj} &\sim  \gamma_{ij}\text{Gamma}\left(\exp\left(\xi_{kj}\right), 1\right) +  (1 - \gamma_{ij} )\delta_{0}\left(\phi_{ij}\right), \mbox{ and} \\
    \gamma_{ij} &\sim \text{Bernoulli}\left(\pi_{\gamma_j}\right), 
\end{split}
\end{align}  
 
where  $\phi_{ij}/\sum_{j'=1}^J \phi_{ij'} $ is the relative abundance of the $j^{th}$, $j=1,\dots,J$, taxon for the $i^{th}$, $i=1,\dots,N$, observation, $\gamma_{ij}$ represents an at-risk indicator, and $\pi_{\gamma_j} \sim \text{Beta}\left(\alpha_{\gamma}, \beta_{\gamma}\right)$ is the probability of an at-risk observation for the $j^{th}$ taxon.    The hyperparameters $\alpha_{\gamma}$ and $\beta_{\gamma}$ control the probability of an at-risk observation, where  $\frac{\alpha_{\gamma}}{\alpha_{\gamma} + \beta_{\gamma}}$ is the expected probability a priori.  For zero counts (i.e., $z_{ij} = 0$), $\gamma_{ij} = 1$ indicates an at-risk zero and $\gamma_{ij} = 0$ a structure zero. Note that $\gamma_{ij} = 1$ for $z_{ij} > 0$. The cluster-specific concentration parameters, $\xi_{kj},$  govern the corresponding relative abundances and are assumed to follow a $\text{Normal}\left( \mu_j, \sigma^{2}\right)$. Given the dimension of the parameter space  found in microbiome applications, we recommend  setting $\mu_j = \log (s*\bar{RA}_j)$, where $s$ is a scaling parameter and $\bar{RA}_j$ is the average relative abundance for the $j^{th}$ taxon observed in the data. Setting $s$ large and $\sigma^{2}$ small will result in the prior for the relative abundances concentrating  around the observed mean. For interpretation, $s$ can be thought of as the hypothetical total number of reads used to inform the prior. 

\subsection{Posterior Sampling and Inference}

\begin{figure}[H] 
\centering
 \includegraphics[width=0.75\textwidth]{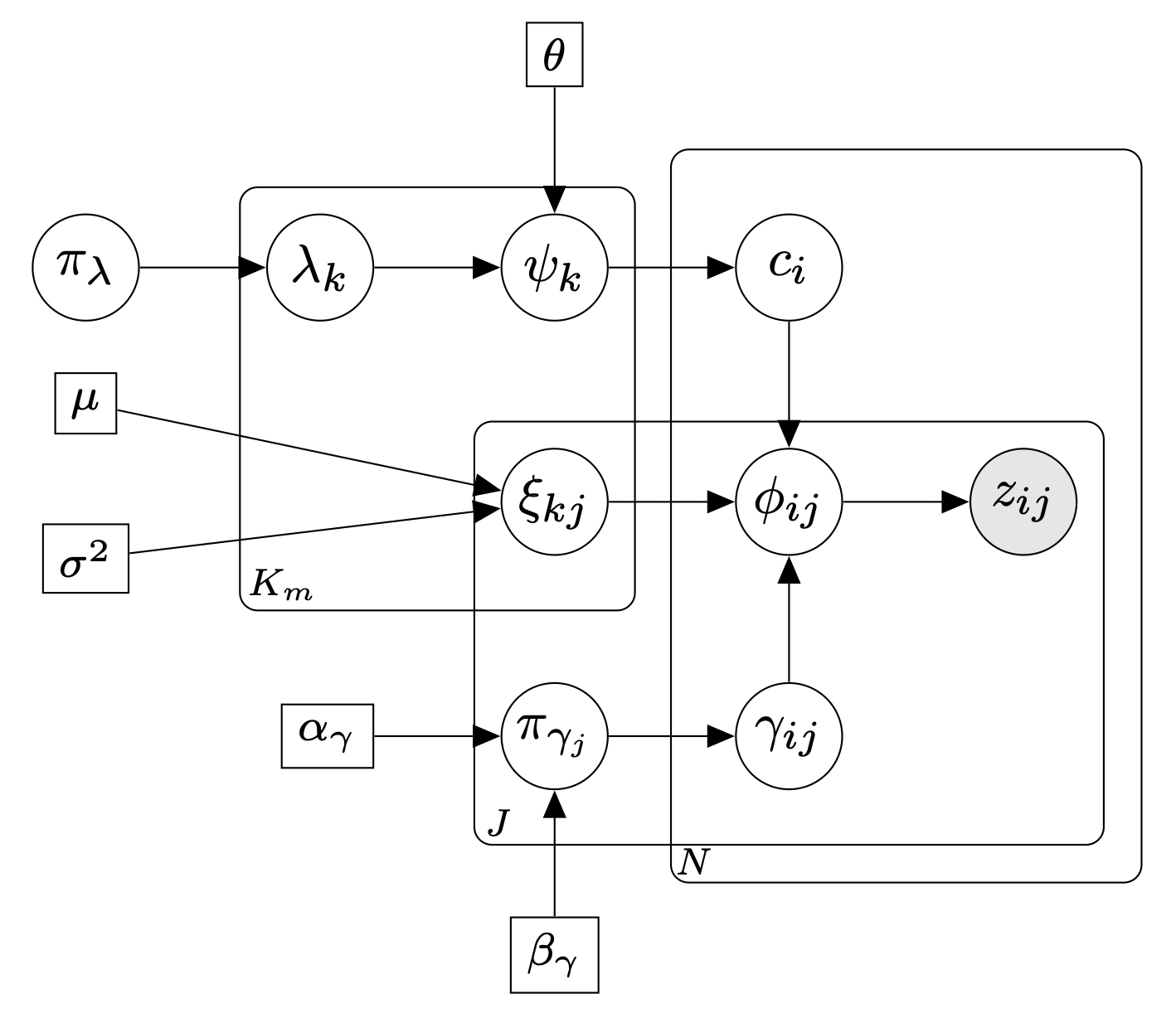} %
\caption{Graphical representation of the proposed clustering framework for zero-inflated multivariate compositional count data. The white (shaded) nodes represent the parameters (observed data). The rectangular (circular) shapes represent fixed (random) variables.}
\label{fig:mod_graphical_full} 
\end{figure}

\noindent The full joint posterior distribution of the proposed DSDM$^3$-ZIDM  is written as 
\begin{align}
    p\left( \bm{\phi}, \bm{c}, \bm{\gamma}, \bm{\xi}, \bm{\pi}_{\gamma}, \bm{\psi}, \bm{\lambda} \mid \bm{z}\right) &\propto p\left(\bm{z},  \bm{\phi}, \bm{c}, \bm{\gamma}, \bm{\xi}, \bm{\pi}_{\gamma}, \bm{\psi}, \bm{\lambda}\right) \nonumber \\
    &= p\left(\bm{z} \mid  \bm{\phi}\right) p\left(\bm{\phi} \mid \bm{c},\bm{\gamma}, \bm{\xi}\right) p\left(\bm{\xi}\right) p\left(\bm{\gamma} \mid \bm{\pi}_{\gamma}\right) p\left(\bm{\pi}_{\gamma}\right) p\left(\bm{c} \mid  \bm{\psi}\right) p\left(\bm{\psi} \mid \bm{\lambda}\right) p\left(\bm{\lambda}\right) , \nonumber
\end{align}
 
\noindent where  $\bm{z}$ represents the $N \times J$ matrix of counts observed across individuals,  $\bm{\phi}$ the $N \times J$ matrix of  individual-specific relative abundances, $\bm{c}$ the $N$-dimensional vector of cluster assignments, $\bm{\gamma}$   the $N \times J$ matrix of at-risk indicators, $\bm{\xi}$ the $K_{m}\times J$ matrix of cluster-specific concentration parameters, $\bm{\psi}$   the $K_{m}$-dimensional vector of potential  cluster weights, $\bm{\pi}_{\gamma}$ the $J$-dimensional vector of at-risk probabilities, and $\bm{\lambda}$   the $K_{m}$-dimensional vector of  indicators for the cluster components. Figure \ref{fig:mod_graphical_full} presents a graphical representation of the proposed model in the context of zero-inflated multivariate  count data.

To generate samples from the posterior distribution, we construct a Metropolis-Hastings within Gibbs algorithm. In practice, we are typically only interested in drawing inference on a subset of the parameters in the model. Therefore, we integrate out $\bm{\phi}$ and  $\pi_{\gamma_j}$ to reduce the computational cost of the resulting MCMC sampler. Following \cite{miller2018mixture}, we could implement Algorithm 8 of \cite{neal2000markov} coupled with a Split-Merge sampler  \citep{jain2007splitting} to update the cluster allocations, $\boldsymbol{c}$. However due to the dimension of the cluster-specific parameters, $\boldsymbol{\xi}_k$, we observed poor mixing for the number of active clusters. Instead, we implemented the telescoping sampler proposed by \cite{fruhwirth2021generalized}, in which the full conditional distribution of $K$, and thus the creation of a new component, does not depend on the cluster parameters. As a result, we observed improved mixing and convergence of the MCMC sampler overall.  More technical details of the MCMC algorithm and relevant derivations are provided in the Supplementary Material. To determine the final cluster assignment for each observation, we applied the \texttt{salso} algorithm \citep{salso}, a randomized greedy search algorithm,  to minimize the lower bound of the variation of information loss \citep{meilua2003comparing}.

\section{Simulation Study}\label{sec::sim}

Before applying the model to   data collected in the EDD study, we first performed a simulation study to evaluate  the clustering performance of the proposed method and compare it to alternative distance- and model-based methods for clustering  multivariate compositional count data in six different scenarios. Specifically, we compared the proposed  DSDM$^3$ to a Dirichlet process (DP) mixture model with a Dirichlet-tree multinomial distribution for the counts \citep{mao2022dirichlet} and a shrinkage-based sparse Dirichlet-multinomial mixture model (sSDM$^{3}$), similar to \cite{saraiva2020bayesian}, with a ZIDM for the counts (DSDM$^3$-ZIDM, DP-DTM, and sSDM$^{3}$-ZIDM,  respectively). Additionally, we compared the model to 
a finite Dirichlet-multinomial mixture model (DM$^3$), similar to \cite{holmes2012dirichlet}, implemented with a ZIDM for the counts (DM$^{3}$-ZIDM). To assess how ignoring zero-inflation may affect inference, we also evaluated the clustering performance of the  DSDM$^3$ and sSDM$^{3}$ with a DM likelihood for the counts (DSDM$^3$-DM and sSDM$^{3}$-DM, respectively). Further, we compared the model to the distance-based clustering method PAM \citep{PAM}, using various distance metrics: Aitchison (PAM-AT)  \citep{aitchison2000logratio}, Bray-Curtis (PAM-BC) \citep{bc_dist}, unweighted UniFrac   (PAM-UU), weighted UniFrac  (PAM-WU)  \citep{unifrac_dist}. Note that DP-DTM, PAM-UU, and PAM-WU require specifying a phylogenetic tree for the microbial count data.  Lastly, we compared to a Gaussian mixture model (GMM) with an additive log-ratio  transformation for the compositional count data \citep{egozcue2003isometric, mclust_package}. 

For the Bayesian mixture models, we assumed similar prior assumptions when applicable (see the Supplementary Material for more details). Each model was run for 10,000  MCMC iterations, treating the first 5,000 iterations as burn-in. Final cluster assignments for the Bayesian models were obtained using the \texttt{salso} algorithm. To determine the number of clusters using the finite mixture models  (i.e., DM$^{3}$-ZIDM and GMM) and distance-based methods (i.e., PAM-AT, PAM-BC, PAM-UU, and PAM-WU), we fit each model with two to ten different components and selected a final model for inference using the Bayesian information criterion (BIC; \citep{swartz2004bayesian}) and average silhouette width, respectively. Clustering performance of the methods was evaluated using the adjusted Rand index (ARI; \citep{hubert1985comparing}), where higher values imply better clustering.


In scenarios 1 - 5, we generated data to evaluate and compare the methods' clustering performance with varying percentages of zero cell counts, numbers of taxa, and numbers of clusters using an approach similar to that taken in \cite{shi2023sparse}. In each scenario, we first specify the number of observations, $N$, the number of taxa which are not differentiated by cluster assignment, $J_{\text{noise}}$, the number of taxa that are cluster-specific, $J_{\text{signal}}$, the total depth for the noise taxa, $\dot{z}_{\text{noise}}$, the total depth for the signal taxa, $\dot{z}_{\text{signal}}$, and the expected proportion of at-risk observations in each data set. In scenarios 1, 2, and 3, we generated two true clusters and 100 taxa with increasing percentages of zero cell counts in the data (i.e., 28\%, 51\%, and  73\%, respectively). The data in scenarios 4 and 5 were simulated similar to scenario 2, however in scenario 4, we increased the number of taxa to 250 and in scenario 5 the number of true clusters to six. For scenarios 1 - 5, the sequencing depth was set to 5,000 reads per sample (4,000 for noise taxa and 1,000 for signal taxa).  Additionally, we compared the models in a setting designed with a more complex correlation structure among the counts (scenario 6). Specifically, multivariate count data were generated using a Dirichlet-tree multinomial distribution with added zero-inflation. The phylogenetic structure  used to simulate these data was obtained from the taxonomy observed in the application study. Additionally, we set the sequencing depth to match the median depth of the EDD dataset (i.e., 2,500 reads per sample). See Table \ref{tab:simulated_data_summarize} and the Supplementary Material for more details of each scenario, including a plot of the phylogenetic tree used to simulate data in scenario 6. 

\vspace{0.25cm}

\begin{table}[H]
\centering
\resizebox{\columnwidth}{!}{%
\begin{tabular}{ccccc}
\hline
Scenario & $K$ & $N$ ($N_1$, \dots, $N_K$ )   & $J$ $ (J_{\text{noise}}, J_{\text{signal}} )$ & Average Proportion of Zero Counts \\ \hline
1        & 2   & 100 (50, 50)                 & 100 (80, 20)                                  & 0.28                       \\
2        & 2   & 100 (50, 50)                 & 100 (80, 20)                                  & 0.51                      \\
3        & 2   & 100 (50, 50)                 & 100 (80, 20)                                  & 0.73                      \\
4        & 2   & 100 (50, 50)                 & 250 (200, 50)                                 & 0.54                      \\
5        & 6   & 150 (30, 30, 20, 20, 25, 25) & 100 (80, 20)                                  & 0.50  \\
6        & 4   & 300 (75, 75, 75, 75)         & 79*                                           & 0.58                        \\ \hline
\end{tabular}
}
\caption{Summary of the Simulation Study Scenarios. Each scenario outlines different configurations for true number of clusters, $K$, sample size, $N$, with $N_k$ representing the number of observations in each true cluster in parentheses, the total number of taxa, $J$, comprising $J_{\text{noise}}$ taxa that were undifferentiated across clusters and $J_{\text{signal}}$ taxa that differentiated the clusters, and the average proportion of zero counts computed over 20 replicate datasets. * - see the Supplementary Material for more details of the data generation in this scenario.}
\label{tab:simulated_data_summarize}
\end{table}

\vspace{0.15cm}

Table \ref{tab:simulated_ari} presents the clustering performance results of the methods in each simulation scenario. Overall, the proposed DSDM$^3$-ZIDM obtained similar or improved clustering performance in terms of the average ARI compared to the alternative distance- and model-based approaches in all six scenarios, with the exception of DM$^3$-ZIDM in scenarios 1 and 2. Recall that DM$^3$-ZIDM is essentially the same model as the proposed DSDM$^3$-ZIDM, however the number of components $K$ are fixed a priori. As the percentage of zero cell counts increased across scenarios 1 - 3, the clustering performance of all methods declined. The proposed method's performance was relatively robust to the number of taxa, $J$. However, we observed a slight decrease in performance in the scenario with six true clusters (i.e., scenario 5). In scenario 6, which was designed to incorporate a more complex correlation structure among the counts, DSDM$^3$-ZIDM and sSDM$^3$-ZIDM outperformed all other methods, including PAM-UU, PAM-WU, and DP-DTM, which were the only methods that incorporated phylogenetic information when performing clustering. In all six scenarios, we observed that versions of the models that accommodated zero-inflation outperformed those that ignored zero-inflation, regardless of the clustering framework used. Among the distance-based and frequentist methods, PAM-WU and PAM-BC demonstrated the best clustering performance. Notably PAM-WU and PAM-BC are equivalent in scenarios 1 - 5, as all simulated samples share the same sequencing depth and no phylogenetic structure is assumed among the OTUs (see the Supplementary Material for more details). In scenario 6, PAM-WU outperformed PAM-BC given the additional information contained in the phylogenetic tree. Lastly, we compared the computation time for each of the methods. DSDM$^3$-ZIDM took roughly 15 minutes in scenarios 1, 2, and 3; 130 minutes in scenario 4; and 45 minutes in scenarios 5 and 6 to run 10,000 MCMC iterations on a  MacBook Pro with an Apple M1 Pro chip (8-core CPU: 6 performance and 2 efficiency cores), 16 GB RAM, and macOS Sequoia 15.4. The alternative Bayesian methods were typically able to run 10,000 iterations quicker than the proposed method in scenarios 1 - 5, especially for those that ignored potential zero-inflation. However, DM$^3$-ZIDM, which relies on model comparison techniques to determine the number of clusters in the data, obtained a cumulative computation time nearly 10 times that of the proposed method. While DP-DTM was the quickest Bayesian method in scenarios 1 - 4, it slowed considerably as the complexity of the data increased in scenarios 5 and 6. The frequentist and distance-based approaches were much faster than the Bayesian methods, as expected.

\begin{table}[H]
\centering
\begin{adjustbox}{width=\textwidth}
\begin{tabular}{ccccccc}
\hline
              & \multicolumn{6}{c}{Scenario}                                                       \\ \cline{2-7} 
Model         & 1           & 2           & 3           & 4            & 5           & 6           \\ \hline
DSDM$^3$-ZIDM & 0.91 (0.09) & 0.84 (0.15) & 0.39 (0.40) & 0.90 (0.11)  & 0.70 (0.13) & 0.86 (0.04) \\
sSDM$^3$-ZIDM & 0.90 (0.31) & 0.70 (0.47) & 0.14 (0.31) & 0.90 (0.23)  & 0.34 (0.15) & 0.86 (0.03) \\
DM$^3$-ZIDM & 1.00 (0.00) & 1.00 (0.02) & 0.36 (0.41) & 0.88 (0.31)  & 0.59 (0.08) & 0.52 (0.26) \\
DSDM$^3$-DM   & 0.05 (0.22) & 0.00 (0.00) & 0.00 (0.00) & 0.00 (0.00)  & 0.00 (0.00) & 0.05 (0.05) \\
sSDM$^3$-DM   & 0.00 (0.00) & 0.00 (0.00) & 0.00 (0.00) & 0.00 (0.00)  & 0.00 (0.00) & 0.03 (0.05) \\
DP-DTM     & 0.00 (0.02) & 0.00 (0.02) & 0.00 (0.01) & 0.00 (0.01)  & 0.01 (0.02) & 0.00 (0.00) \\
PAM-BC      & 0.76 (0.18) & 0.42 (0.21) & 0.26 (0.15) & 0.38 (0.27)  & 0.43 (0.08) & 0.06 (0.08) \\
PAM-AT      & 0.01 (0.02) & 0.01 (0.02) & 0.07 (0.09) & 0.02 (0.05)  & 0.03 (0.02) & 0.01 (0.01) \\
PAM-UU      & 0.01 (0.04) & 0.01 (0.02) & 0.02 (0.03) & 0.00 (0.01)  & 0.03 (0.02) & 0.04 (0.03) \\
PAM-WU      & 0.76 (0.18) & 0.42 (0.21) & 0.26 (0.15) & 0.39 (0.26)  & 0.43 (0.08) & 0.37 (0.06) \\
GMM         & 0.00 (0.03) & 0.01 (0.03) & 0.00 (0.02) & -0.01 (0.02) & 0.01 (0.02) & 0.01 (0.01) \\ \hline
\end{tabular}
\end{adjustbox}
\caption{Simulation Results: Average Adjusted Rand Index (ARI) with standard deviations in parentheses for all methods across 20 replicate data sets. Higher values of ARI represent better clustering performance.}
\label{tab:simulated_ari}
\end{table} 

\begin{table}[H]
\centering
\begin{adjustbox}{width=\textwidth}
\begin{tabular}{ccccccc}
\hline
              & \multicolumn{6}{c}{Scenario}                                                               \\ \cline{2-7} 
Model         & 1            & 2            & 3            & 4             & 5            & 6              \\ \hline
DSDM$^3$-ZIDM & 16.52 (1.17) & 12.61 (0.71) & 14.54 (0.62) & 130.85 (9.79) & 44.79 (3.04) & 44.20 (2.19)   \\
sSDM$^3$-ZIDM & 16.51 (1.12) & 13.19 (1.05) & 32.65 (2.93) & 132.67 (9.53) & 44.47 (3.34) & 84.14 (5.58)   \\
DM$^3$-ZIDM & 153.80 (1.79) & 116.55 (1.07) & 134.63 (1.48) & 1217.04 (19.81) & 358.83 (11.69) & 368.57 (1.87) \\
DSDM$^3$-DM   & 13.07 (0.91) & 8.90 (0.61)  & 8.67 (0.38)  & 89.16 (5.51)  & 32.67 (1.68) & 25.98 (1.87)   \\
sSDM$^3$-DM   & 13.51 (0.97) & 9.11 (0.71)  & 9.01 (0.37)  & 89.68 (6.25)  & 31.44 (1.37) & 51.50 (3.82)   \\
DP-DTM     & 10.21 (2.83) & 7.07 (1.40)  & 6.16 (1.22)  & 11.86 (3.69)  & 65.14 (9.45) & 252.58 (25.48) \\
PAM-BC      & 0.00 (0.00)   & 0.00 (0.00)   & 0.00 (0.00)   & 0.00 (0.00)     & 0.00 (0.00)    & 0.01 (0.01)   \\
PAM-AT      & 0.00 (0.00)   & 0.00 (0.00)   & 0.00 (0.00)   & 0.00 (0.00)     & 0.00 (0.00)    & 0.01 (0.01)   \\
PAM-UU      & 0.04 (0.02)   & 0.04 (0.02)   & 0.04 (0.02)   & 0.12 (0.02)     & 0.25 (0.01)    & 0.83 (0.09)   \\
PAM-WU      & 0.04 (0.01)   & 0.04 (0.01)   & 0.04 (0.02)   & 0.12 (0.02)     & 0.27 (0.06)    & 0.82 (0.09)   \\
GMM         & 0.00 (0.00)   & 0.00 (0.00)   & 0.01 (0.00)   & 0.01 (0.01)     & 0.04 (0.01)    & 0.07 (0.01)   \\ \hline
\end{tabular}
\end{adjustbox}
\caption{Simulation Results: Average computation time in minutes with standard deviations in parentheses for all methods across 20 replicate data sets. For models that rely on model comparison techniques to determine the number of clusters, we report the cumulative computation time for each model fit.}
\label{tab:simulated_time}
\end{table}

\vspace{ 0.5cm }

\section{Zero-Inflated Microbiome Data Application } \label{sec::app_analysis}

In this section, we apply our proposed clustering method to  microbiome data collected in a study investigating the relation between gut microbial composition and enteric diarrheal disease (EDD; \citep{singh2015intestinal}).  
These data were made available by \cite{duvallet2017meta} as part of a meta-analysis for gut microbiome studies. Following the data processing pipeline described in \cite{duvallet2017meta}, the data investigated in this analysis  consisted of $N=303$ observations (221 EDD patients and 82 healthy controls) of $J=79$ taxa at the genus level with 58\% zero counts.

For the proposed DSDM$^{3}$-ZIDM, we set $K_m  = 10$, $\theta = 0.1$,  $\alpha_{\gamma} = 1$, $\beta_{\gamma} = 1$, $\pi_{\lambda} = 0.5$, $s = 200$, $\sigma^{2} = 10$, and $\sigma^{2}_{\text{MH}} = 1$. Values of $\sigma^{2}_{\text{MH}}$ and $s$ were chosen to ensure adequate mixing of the MCMC sampler. The MCMC sampler was initialized at a singleton cluster with $ \xi_{1j} = \log( 200 * \bar{RA}_j)$ and run for 15,000 iterations.  We treated the first 5,000 iterations as burn-in and used the remaining 10,000 samples for inference. Convergence was then assessed visually using  trace plots for  the mixture weights, $\bm{w}$, and cluster-specific concentration parameters for the relative abundances, $\boldsymbol{\xi}_{k}$.  See the Supplementary Material for more details. Cluster allocation was then determined using the \texttt{salso} method to minimize the lower bound of the variation of information loss. 

\begin{figure}[H]
 \centering
\includegraphics[width=0.95\textwidth]{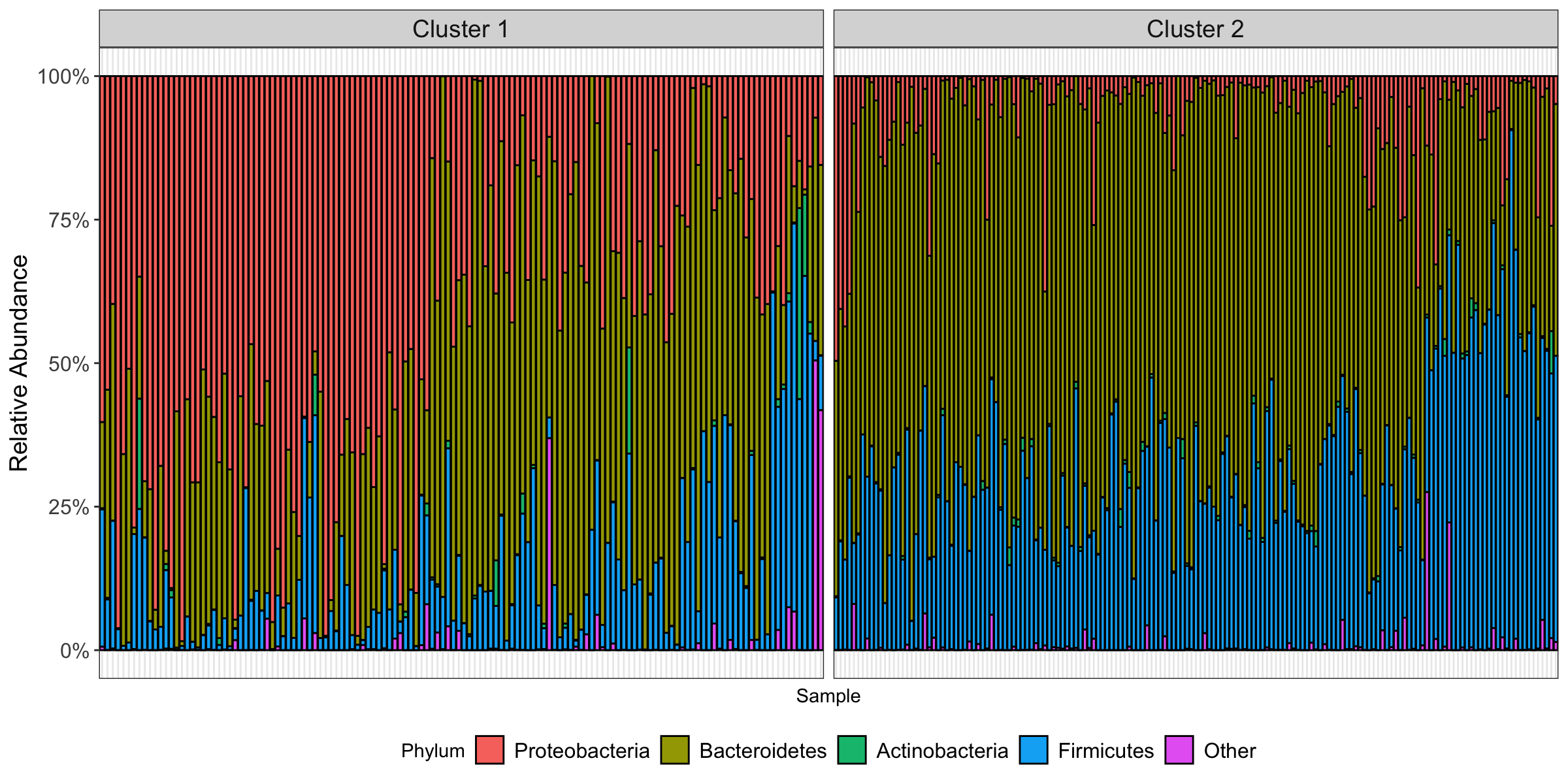}
 \caption{EDD Application Results: Observed relative abundances of each patient in the two main clusters obtained with DSDM$^3$-ZIDM. Relative abundances are aggregated at the phylum level for ease of presentation.} \label{fig:rela_phyla_2clus} 
\end{figure}

\begin{figure}[H]
\centering
    \includegraphics[width=0.75\textwidth]{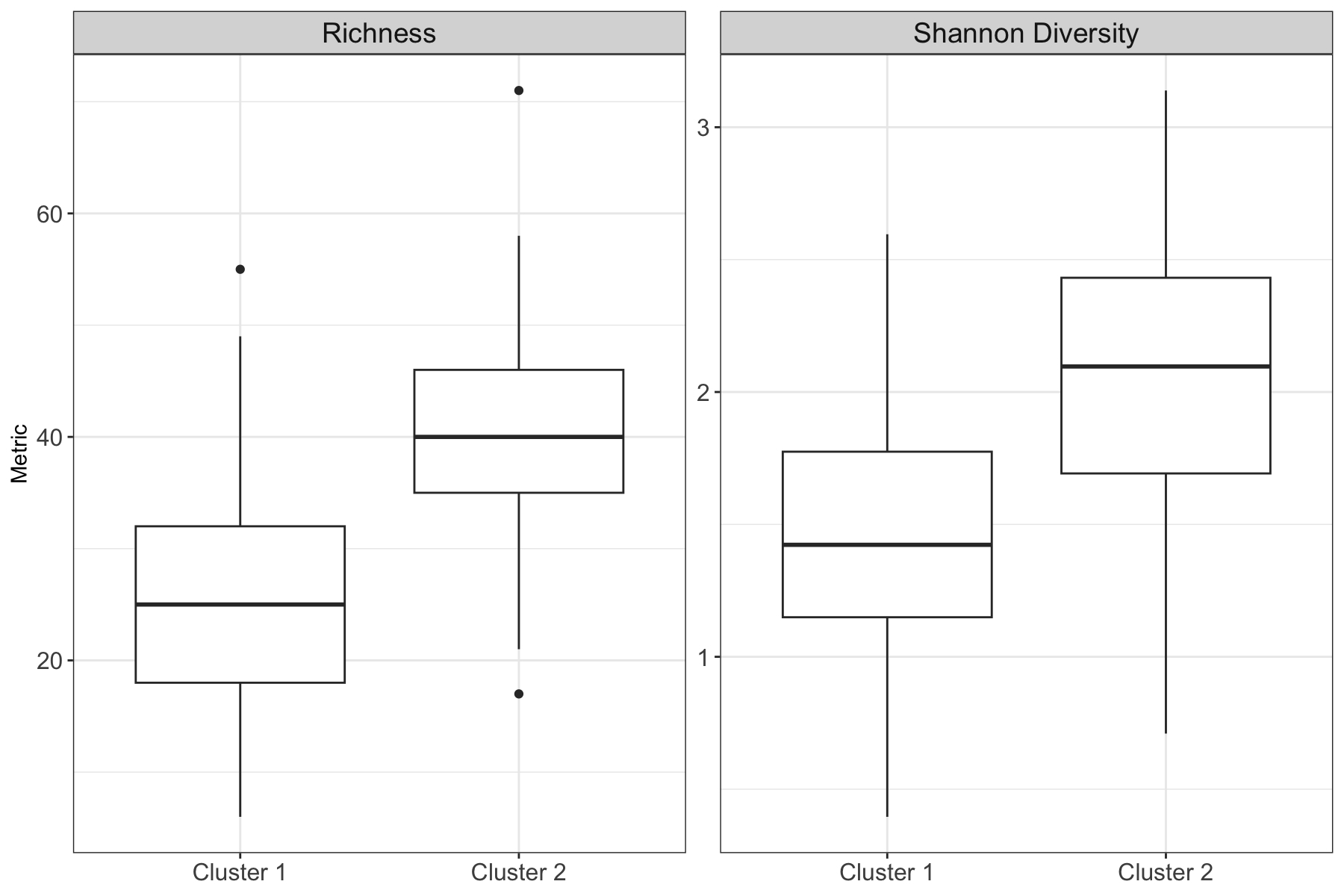}
\caption{EDD Application Results: Boxplots of the richness and diversity of microbial communities across the two main clusters obtained with DSDM$^3$-ZIDM. The left panel depicts the richness (number of different species) for each cluster, while the right panel illustrates the Shannon diversity index (a measure of species diversity accounting for both abundance and evenness).}
\label{fig:alpha_div_2clus}
\end{figure}

In a recent study assessing gut microbial community composition conducted on three large metagenomic datasets, the authors identified three enterotypes: one characterized by $Bacteroides$, another by $Prevotella$, and third by Firmicutes with $Ruminococcus$ typically the most abundant \citep{costea2018enterotypes}. Our analysis identified two main  clusters in the data; one with  genera balanced between phyla Firmicutes ($0.40$), Proteobacteria ($0.28$),  and Bacteroidetes ($0.27$) (cluster 1) and the other by  Bacteroidetes ($0.47$) and Firmicutes ($0.42$) (cluster 2). Posterior estimates of the relative abundances obtained with  DSDM$^3$-ZIDM  are  in parentheses for reference. At the genus level, cluster 1 was dominated by a combination of $Bacteroides$ ($0.15$) and $Cronobacter$ ($0.13$), while cluster 2 was dominated by $Bacteroides$ ($0.33$). Figure \ref{fig:rela_phyla_2clus} presents the observed relative abundances aggregated at the phyla level for ease of presentation for the two main clusters. In the Supplementary Material, we additionally present the abundances aggregated at the family level for reference.  Of the 136  individuals assigned to cluster 1, 132 were EDD patients and 4 were healthy controls. Whereas almost half of the 163 individuals assigned to cluster 2 were healthy controls (78/163). We observed that richness and Shannon diversity indices were lower for the cluster dominated by EDD patients (Figure \ref{fig:alpha_div_2clus}), similar to the results presented in \cite{singh2015intestinal}. Additionally, the model identified 2 singleton clusters and another cluster of only two individuals. Interestingly, the individuals assigned to these clusters were all EDD patients with microbial compositions dominated by $Cronobacter$ ($>0.86$ in each). Recall that we also observed cluster 1, which was mostly EDD patients, having higher levels of genus $Cronobacter$. $Cronobacter$ is a gram-negative bacterium  associated with foodborne diseases and infections \citep{forsythe2018updates}. Previously, researchers have found $Cronobacter$ levels increased for patients with colorectal polyps, colorectal cancer, irritable bowel syndrome, autism spectrum disorder, and metabolic syndrome  \citep{wu2025characteristics}.

\subsection{Sensitivity Analyses} 

In this section, we perform a thorough sensitivity analysis of the application results to hyperparameter specification for DSDM$^3$-ZIDM. Thereafter, we compare the results obtained in the application study to those obtained with alternative distance- and model-based methods. To assess the model’s sensitivity to prior specification, we set each of the hyperparameters to default values (i.e., those used in application study) and then evaluated the effect of manipulating each term on inference. We observed that the clustering results were relatively sensitive to   $\theta$, which can be interpreted as the concentration parameter of the Dirichlet distribution  on the mixture weights (see Table \ref{tab::sensitivity}). While the model was robust to smaller values of $\theta$, larger values resulted in the MCMC chain getting stuck at $K=1$ mixture components. These results were not surprising as previous studies have shown that clustering performance is sensitive to the prior specification of the mixture weights \citep{celeux2019model, fruhwirth2019here}. We found that inference was relatively robust to the variance of the proposal distribution for $\boldsymbol{\xi}$. However, the model was moderately sensitive to $s$, the scaling factor for  the prior mean of $\boldsymbol{\xi}$. Specifically, we found that the model had difficulty finding favorable new cluster parameters given their high dimensionality, which resulted in poorer mixing of the MCMC sampler. We observed that reducing the probability of an at-risk observation resulted in an increase in the number of smaller clusters. Lastly, as the prior probability for the number of active components decreased, so did the number of clusters, as expected.

\vspace{0.5cm}

\begin{table}[H]
\centering
\begin{tabular}{cccccccccc}
\hline
         & $\theta = 0.01$ &  & $\theta = 1$         &  & $\sigma_{MH}^2 = 0.1$ &  & $\sigma_{MH}^2 = 10$  &  & $s = 100$             \\ \cline{2-2} \cline{4-4} \cline{6-6} \cline{8-8} \cline{10-10} 
Clusters & 3               &  & 1                    &  & 4                     &  & 6                     &  & 4                     \\
ARI      & 0.98            &  & 0.00                 &  & 0.96                  &  & 0.95                  &  & 0.65                  \\
\cline{2-10}
         & $s=300$         &  & $\beta_{\gamma} = 4$ &  & $\beta_{\gamma} = 9$  &  & $\pi_{\lambda} = 0.2$ &  & $\pi_{\lambda} = 0.1$ \\ \cline{2-2} \cline{4-4} \cline{6-6} \cline{8-8} \cline{10-10} 
Clusters & 5               &  & 9                    &  & 5                     &  & 4                     &  & 1                     \\
ARI      & 0.95            &  & 0.48                 &  & 0.44                  &  & 0.97                  &  & 0.00                  \\ \hline
\end{tabular}
\vspace{1cm}
\caption{ Sensitivity Analysis for EDD Application Results: Sensitivity of the results obtained with DSDM$^3$-ZIDM to hyperparameter specification. ARI - Adjusted Rand index. Clusters - the number of clusters estimated by the model. \label{tab::sensitivity}}
\end{table}

 To assess the sensitivity of the application results to model selection and specification, we first reanalyzed the data with alternative MFM models that place different priors on the number of components in the model. Specifically, we fit a MFM model assuming $K$ followed a truncated Poisson (MFM-P), geometric (MFM-G), and beta-negative-binomial (MFM-BNB) distribution with a ZIDM for the count data. The hyperparameters for each model were specified so that the prior mean for the number of components matched DSDM$^3$-ZIDM. All other hyperparameters were set similar to the application study. Posterior samples were obtained using the telescoping sampler of \cite{fruhwirth2021generalized}, and post-hoc inference was performed with the \texttt{salso} algorithm. The alternative MFM models all identified two main clusters with comparable numbers of healthy individuals and EDD patients and similar microbial compositions, resulting in all methods obtaining $>0.95$ ARI compared to the cluster allocation found with the proposed method. Additionally, the MFM-P, MFM-G, and MFM-BNB models identified one, two, and three singleton clusters of EDD patients with high levels of $Cronobacter$, respectively. Similar cluster allocations were found using the DM$^3$-ZIDM model. The DP-DTM model was not able identify any clusters in the data. sSDM$^3$-DM and sSDM$^3$-ZIDM  suggested 8 and 9 clusters with 0.29 and 0.39 ARI compared to the cluster allocation found with the proposed method, respectively. For comparison, we also fit each of the PAM methods using the distance metrics evaluated in the simulation study with two to ten clusters and then selected the model with higher average silhouette width for inference.  While each of the distance-based methods identified two clusters in the data (i.e., one dominated by $Cronobacter$ and $Bacteroides$ and another by $Bacteroides$), the results differed from  those found with the proposed method  (i.e., ARIs ranging from $0.34$ to $0.53$). Lastly, we evaluated the sensitivity of the results to likelihood assumptions by applying the DSDM$^3$ with a DM distribution for the multivariate count data (DSDM$^3$-DM), which ignores potential zero-inflation. All else equal, DSDM$^3$-DM was unable to identify any clusters in the data,  highlighting the importance of accommodating zero-inflation when clustering microbiome data. 

 \section{Discussion}
  
In this work, we performed a cluster analysis on microbial composition data collected in a study designed to investigate the relation between gut microbial composition and enteric diarrheal disease using a novel Bayesian semiparametric clustering method for multivariate compositional count data with zero-inflation.  Through sensitivity analysis, we show how inference can depend heavily on  decisions regarding the clustering method and  its specification, underscoring some of the challenges of clustering microbiome data that are commonly faced by the field \citep{costea2018enterotypes}. Our approach uses a discrete sparse mixture modeling framework to determine cluster allocation, which allows the mixture weights of empty cluster components to take on zero values. As such, it can be seen as an extension to the work of \cite{koslovsky2023bayesian} that additionally accounts for heterogeneity in zero-inflated multivariate count data, effectively using a ZIDM model to accommodate zero-inflation in the count data as well as sparsity in the mixture weights. Additionally, we show how the proposed clustering method belongs to the class of MFM models. This connection opens up a suite of sampling algorithms for posterior inference, as demonstrated in \cite{miller2018mixture} and \cite{fruhwirth2021generalized}. In the Supplementary Material, we provide derivations for the MCMC algorithm using the telescoping sampler and those needed to implement our approach with the algorithms described in \cite{miller2018mixture} and \cite{neal2000markov}. In the simulation study, we find that our model is able to obtain similar or improved  clustering performance compared to alternative distance- and model-based clustering methods while properly accounting for zero-inflation in the data. 

In the EDD application study, we cluster the  data  at the genus level. However in practice, the model is agnostic to the level in which the multivariate count data are aggregated prior to analysis. While designed for zero-inflated multivariate compositional count data collected in human microbiome research settings, the semiparametric clustering framework is flexible to other data structures by adjusting the likelihood function accordingly. Another future extension of the proposed model we aim to explore is how different prior specifications for the active component indicators, $\lambda_k$, or hyperpriors for $\theta$, which control the mixing weights, may induce more desirable clustering behavior and/or accommodate available information to improve clustering performance and inference. For example, we may allow the mixture weights to depend on covariate information so that individuals with similar covariate values are more likely to cluster together. Further, the notion of placing a point mass at zero for mixture weights was also recently proposed in infinite settings via the atom-skipping process and the plaid atoms model for multiple groups \citep{bi2023class}. Future work could explore the use of the proposed DSDM$^3$ in grouped settings as a semiparametric alternative.

\section{Acknowledgments}
SK and MDK gratefully acknowledge the support of NSF grant DMS-2245492. The opinions, findings, and conclusions expressed are those of the authors and do not necessarily reflect the views of the NSF.


\section{Supplementary Material}
Mathematical derivations, additional technical details of the MCMC sampler, the data generation process used in the simulation study, and supplementary figures are provided in the Supplementary Materials. The data used in the application study, the code for generating the simulated data, and the code for implementing the proposed method are available in the accompanying \texttt{R} package, \texttt{DSDM$^3$}, which can be accessed from the author’s \href{https://github.com/skorsu/DSDM3/tree/main}{GitHub} repository.

\section*{Data Availability}
The case study data from \cite{singh2015intestinal} are available in the MicrobiomeHD database \citep{duvallet2017meta}(\href{https://zenodo.org/records/569601}{https://zenodo.org/records/569601}).

\bibliographystyle{abbrvnat}
\bibliography{ref}

\end{document}